# *J*-matrix method of scattering for inverse-square singular potential with supercritical coupling I. Theory

Abdulaziz D. Alhaidari[a], Hocine Bahlouli[b], Carlos P. Aparicio[c], and Saeed M. Al-Marzoug[b]

[a] *Saudi Center for Theoretical Physics, P. O. Box 32741, Jeddah 21438, Saudi Arabia*
[b] *Physics Department, King Fahd University of Petroleum & Minerals, Dhahran 31261, Saudi Arabia*
[c] *Alameda de San Antón 45 3°I, CP 30205 Cartagena, Murcia, Spain*

**Abstract**: The *J*-matrix method of scattering was developed to handle regular short-range potentials with applications in atomic, nuclear and molecular physics. Its accuracy, stability, and convergence properties compare favorably with other successful scattering methods. It is an algebraic method, which is built on the utilization of orthogonal polynomials that satisfy three-term recursion relations and on the manipulation of tridiagonal matrices. Recently, we extended the method to the treatment of $r^{-2}$ singular short-range potentials but confined ourselves to the sub-critical coupling regime where the coupling parameter strength of the $r^{-2}$ singularity is greater than $-1/8$. In this work, we expand our study to include the supercritical coupling in which the coupling parameter strength is less than $-1/8$. However, to accomplish that we had to extend the formulation of the method to objects that satisfy five-term recursion relations and matrices that are penta-diagonal. It is remarkable that we could develop the theory without regularization or self-adjoint extension, which are normally needed in the treatment of such highly singular potentials. Nonetheless, we had to pay the price by extending the formulation of the method into this larger representation and by coping with slower than usual convergence. In a follow-up study, we intend to apply the method to obtain scattering information for realistic potential models.

*To the memory of our departed friend, colleague and collaborator Mohammed S. Abdelmonem.*

## I. INTRODUCTION

The *J*-matrix is an algebraic method of scattering that was developed by Heller and Yamani more than 45 years ago to handle quantum scattering with short-range regular potentials [1-3]. It is the function-space analog of the *R*-matrix method of scattering in configuration space [4]. The shortness of the range of the scattering potential in the *R*-matrix method is exemplified by assuming that $V(r) = 0$ for $r \geq R$, where *R* is some finite range parameter. This approximation corresponds to the finiteness of the size of the matrix representation of the potential in the *J*-matrix basis (i.e., $V_{nm} = 0$ for $n,m \geq N$ with *N* being large enough integer). Originally, the method was considered a purely physical tool but then Ismail and his collaborators turned it into a mathematical technique (see, for example, recent studies in Refs. [5-8]). The method is built on the analytic power of orthogonal polynomials in the treatment of the reference problem where $V(r) = 0$. This is then augmented by highly accurate numerical schemes to handle the contribution coming from the finite potential matrix. In such schemes, one employs powerful numerical techniques that include Gauss quadrature, continued fractions, tridiagonal matrices, etc. The viability and accuracy of the method compares favorably with other successful scattering methods. Several research groups use the method regularly to solve atomic, molecular and nuclear physics problems. Most of these groups are in Russia, Saudi Arabia, United States, and Belgium. For an expose of recent developments and selected applications of the method, one may refer to the collected sample work by some of these groups in Ref. [9]. Now, one of the most interesting and highly non-trivial problems in scattering theory is associated with short-range potentials that are singular at the origin. Recently, we extended the *J*-matrix method to the treatment of singular short-range potentials with $r^{-1}$ singularity [10]. The trick was in transferring the singularity from the scattering potential to the reference Hamiltonian forming a new Coulomb-like potential where it is treated exactly using the analytic tools of the method.

Lately, we used the same approach to extend the method to singular short-range potentials with inverse square singularity [11]. This was also accomplished by removing the singularity from the scattering potential and adding it to the kinetic energy operator producing a new orbital term with an *effective* angular momentum that depends on the strength of the inverse square singularity of the potential. Generally, this effective angular momentum is no longer an integer and, in fact, could become complex. The new orbital term makes the reference problem an inverse square singular potential problem, which is known for its quantum anomalies [12-14]. To avoid these complications, we limited our previous study in [11] to the case where $(\ell+\tfrac{1}{2})^2 > A$ with $-A$ being the strength parameter of the inverse square singularity. This situation corresponds to the subcritical coupling regime of the inverse square potential. Moreover, it turned out that this case is a generalized version of the conventional *J*-matrix method with pure kinetic energy reference Hamiltonian but with a non-integer real angular momentum. Due to this fact, the orthogonal polynomials that appear in the solution of the reference problem differ from those in the conventional *J*-matrix.

In this paper, we complement the work in [11] by considering the supercritical coupling regime where $(\ell+\tfrac{1}{2})^2 < A$. However, we encountered an unsurmountable problem, which is the nonexistence of a matrix representation for the reference wave operator that is both tridiagonal and Hermitian at the same time. Consequently, we had to relax the tridiagonal constraint in an effort to preserve Hermiticity of the Hamiltonian, wave operator, and Green's function of the problem. In fact, such difficulty should have been anticipate due to the well-known anomalies in the corresponding inverse square potential with supercritical coupling [12-14]. Most prominent of these anomalies is the rapid and unbounded increase in oscillations of the particle's wavefunction near the origin. Due to this unphysical divergence in oscillations, we proposed in [15] that the interaction gets screened at short distances making the bare coupling parameter $\ell(\ell+1)-A$ acquire an effective renormalized value that falls within the weak range $[0,\tfrac{1}{4}]$. This prevents the oscillations form growing without limit giving a lower bound to the energy spectrum and forcing the reference Hamiltonian of the system to be self-adjoint. Technically, this translates into a regularization scheme whereby the inverse square potential is replaced near the origin by another potential that has the same singularity but with a weak coupling strength. The details of this investigation are found in Ref. [15].

In the following section, we show how to handle this anomalous inverse square problem by rebuilding the formulation of the *J*-matrix method on objects that satisfy five-term recursion relations instead of orthogonal polynomials that satisfy three-term recursions.

## II. REFERENCE SOLUTION AND THE FIVE-TERM RECURSION RELATION

In the atomic units $\hbar = m = 1$, the total Hamiltonian of the problem is $H = -\frac{1}{2}\frac{d^2}{dr^2} + \frac{\ell(\ell+1)}{2r^2} + V(r)$, where $V(r) = -\frac{A/2}{r^2} + U(r)$ and $A$ is a real dimensionless positive parameter. The potential component $U(r)$ is nowhere singular[†] and its range is short. We write $H = H_0 + U(r)$, where $H_0 = -\frac{1}{2}\frac{d^2}{dr^2} + \frac{\ell(\ell+1)-A}{2r^2}$ and the reference wave equation becomes

$$\left[\frac{d^2}{dr^2} - \frac{\ell(\ell+1)-A}{r^2} + k^2\right]\psi(r) = 0, \tag{1}$$

where $k^2 = 2E > 0$. If we let $y = kr$ and write $\psi(r) = y^\gamma f(y)$ then we obtain

$$\left[\frac{d^2}{dy^2} + \frac{2\gamma}{y}\frac{d}{dy} - \frac{\ell(\ell+1)-A+\gamma(1-\gamma)}{y^2} + 1\right]f(y) = 0. \tag{2}$$

This is the differential equation for the Bessel function $J_\nu(y)$ provided that $\gamma = \frac{1}{2}$ and $\nu = \sqrt{(\ell+1/2)^2 - A}$. That is,

$$\left[\frac{d^2}{dy^2} + \frac{1}{y}\frac{d}{dy} - \frac{\nu^2}{y^2} + 1\right]J_\nu(y) = 0. \tag{3}$$

By studying the solution space of Eq. (1), we conclude the following:
1. The two independent solutions of (1) are $\sqrt{y}J_\nu(y)$ and $\sqrt{y}J_{-\nu}(y)$.
2. If $(\ell+\frac{1}{2})^2 > A$ then $\nu$ is real and $\sqrt{y}J_{-\nu}(y)$ diverges at the origin while $\sqrt{y}J_\nu(y)$ is regular everywhere. The two independent solutions $\sqrt{y}J_{\pm\nu}(y)$ are asymptotically sinusoidal with the same amplitude of $\sqrt{2/\pi}$ but with a constant phase difference. This phase angle can be made equal to $\frac{\pi}{2}$ by writing the irregular solution as a special linear combination of $J_{\pm\nu}(y)$ as follows: $\sqrt{y}Y_\nu(y) = \sqrt{y}\left[\cos(\nu\pi)J_{+\nu}(y) - J_{-\nu}(y)\right]/\sin(\nu\pi)$.
3. If $(\ell+\frac{1}{2})^2 < A$ then $\nu$ is pure imaginary and the two independent solutions are regular everywhere. Therefore, the general solution of (1) for all regions of configuration space is a linear combination of $\sqrt{y}J_{\pm\nu}(y)$. Moreover, $J_{-\nu}(y)$ is the complex conjugate of $J_\nu(y)$. Now, the real and imaginary parts of $\sqrt{y}J_\nu(y)$ are asymptotically sinusoidal with exactly $\frac{\pi}{2}$ phase difference. However, their amplitudes are not the same; with the ratio (imaginary/real) equal to $\tanh\left(|\nu|\frac{\pi}{2}\right)$.

---

[†] In general, we can also include 1/r singular component in the potential $U(r)$. That is, we can allow it to take the form $U(r) = \frac{Z}{r} + W(r)$ with $W(r)$ being non-singular and short-range. In that case, we can apply the extended J-matrix method with 1/r singular potentials that we have already developed in [10].

The first case where $(\ell+\frac{1}{2})^2 > A$, is a generalization of the conventional J-matrix with $\ell$ becoming non-integer; specifically, $\ell \to \nu - \frac{1}{2}$. This case was treated in our earlier work in [11]. However, the second case where $(\ell+\frac{1}{2})^2 < A$ is more complicated due to the supercritical coupling that introduces anomalous behavior as noted in the introduction section above. Here, we study this second case and write the two independent solutions of the reference wave equation (1) as linear combinations of $\sqrt{y} J_{\pm\nu}(y)$ such that their asymptotics match the physical requirements, which reads $\sqrt{\frac{2}{\pi}} e^{\pm i(kr+\varphi)}$ where $\varphi$ is some constant global phase. These are written in terms of the two Hankel functions as follows

$$\chi_\pm(r) = A_\pm \sqrt{kr} H^\pm_{i\nu}(kr), \tag{4}$$

where $H^\pm_\alpha(x) = J_\alpha(x) \pm i Y_\alpha(x)$, $A_\pm$ are arbitrary normalization constants and we rewrote $\nu$ as $i\nu$ with $\nu = \sqrt{A - (\ell+1/2)^2}$, which is real and positive. Using the asymptotic formula of the Hankel functions, which reads $\lim_{y \to \infty} \sqrt{y} H^\pm_{i\nu}(y) = \sqrt{\frac{2}{\pi}} e^{\pm\pi\nu/2} e^{\pm i(y-\pi/4)}$ [16,17], we obtain $A_\pm = e^{\mp\pi\nu/2}$ and $\varphi = -\pi/4$.

At this point in the J-matrix method, we choose a complete set of discrete square integrable basis $\{\phi_n(r)\}$ to expand the two independent solutions of the reference problem as $\chi_\pm(r) = \sum_n F^\pm_n(k) \phi_n(r)$. We require that the matrix representation of the reference wave operator in this basis, $\langle\phi_n|\mathcal{J}|\phi_m\rangle$, be both Hermitian and tridiagonal, where $\mathcal{J} = H_0 - E$. Thus, the reference wave equation becomes a three-term recursion relation for the expansion coefficients $\{F^\pm_n\}$ to be solved in terms of orthogonal polynomials in the energy. Now, the expansion series should be identical to the regular solution of the reference wave equation everywhere. On the other hand, it is required to match the irregular solution only asymptotically. In the present problem, however, the two independent solutions of the reference wave equation are regular everywhere. Thus, the series representations should be exactly equal to $\chi_\pm(r)$. Typically for problems with reference wave equations similar to Eq. (1), we arrive at two types of discrete basis sets that are constructed using the Laguerre polynomials $L^\beta_n(z)$, where $\beta$ is a real parameter. One is referred to as the "Laguerre basis" in which the argument of the Laguerre polynomial is $z = \lambda r$, where $\lambda$ is a real positive scale parameter. The other is the "oscillator basis" in which the argument is $z = (\lambda r)^2$. We start by adopting the Laguerre basis and leave the treatment of the oscillator basis to Appendix A. Thus, we choose the following parameterized basis elements

$$\phi_n(\lambda r) = \sqrt{\frac{\Gamma(n+1)}{\Gamma(n+\beta+1)}} e^{-\lambda r/2} (\lambda r)^\alpha L^\beta_n(\lambda r), \tag{5}$$

where the parameters $\alpha$ and $\beta$ are real with $\beta > -1$. We expand the two reference wavefunctions in this basis as follows

$$A_\pm \sqrt{\mu x} H_{i\nu}^\pm (\mu x) = \sum_{n=0}^\infty F_n^\pm (k) \phi_n (x), \qquad (6)$$

where $x = \lambda r$ and $\mu = k/\lambda$. Using the orthogonality of the Laguerre polynomials, we obtain the following integral representation of the expansion coefficients $F_n^\pm (k)$

$$F_n^\pm (k) = A_\pm \sqrt{\frac{\mu \Gamma(n+1)}{\Gamma(n+\beta+1)}} \int_0^\infty H_{i\nu}^\pm (\mu x) e^{-x/2} x^{\beta-\alpha+\frac{1}{2}} L_n^\beta (x) dx . \qquad (7)$$

One can show that the complex conjugate of $A_+ H_{i\nu}^+ (\mu x)$ is $A_- H_{i\nu}^- (\mu x)$. Since $\phi_n (x)$ is real then $F_n^\pm (k)$ are complex conjugates of each other and we can write

$$F_n^\pm (k) = C_n (k) \pm i S_n (k), \qquad (8)$$

where $\{C_n, S_n\}$ are real functions of the energy. Applying the reference wave operator (1), which reads $\mathcal{J} = H_0 - E = -\frac{\lambda^2}{2}\left[\frac{d^2}{dx^2} + \frac{\nu^2 + \frac{1}{4}}{x^2} + \mu^2\right]$, on the basis element (5) and using the differential equation of the Laguerre polynomial and its differential property, $x\frac{d}{dx}L_n^\beta = nL_n^\beta - (n+\beta)L_{n-1}^\beta$, we obtain

$$\mathcal{J}\phi_n (x) = -\frac{\lambda^2}{2} e^{-x/2} x^{\alpha-2} \sqrt{\frac{\Gamma(n+1)}{\Gamma(n+\beta+1)}}$$
$$\left\{\left[\nu^2 + \left(\alpha - \tfrac{1}{2}\right)^2 + n(2\alpha - \beta - 1) - x(n+\alpha) + x^2\left(\mu^2 + \tfrac{1}{4}\right)\right] L_n^\beta (x) - (2\alpha - \beta - 1)(n+\beta) L_{n-1}^\beta (x)\right\} \qquad (9)$$

If we use the recursion relation of the Laguerre polynomial inside the square brackets of this equation then we obtain the corresponding matrix representation of the reference wave operator $\mathcal{J}(k)$. Since $x L_n^\beta (x)$ gives three terms proportional to $L_n^\beta$ and $L_{n\pm1}^\beta$, then reiteration will force $x^2 L_n^\beta (x)$ to produce five terms proportional to $L_n^\beta$, $L_{n\pm1}^\beta$ and $L_{n\pm2}^\beta$. To restrict to tridiagonal matrix representation for $\mathcal{J}(k)$ we must then impose one of the following two conditions:

1. $2\alpha = \beta + 2$ and $\mu^2 = -\tfrac{1}{4}$, or
2. $2\alpha = \beta + 1$ and $\nu^2 = -\tfrac{1}{4}\beta^2$.

However, reality of the reference Hamiltonian $H_0$ does not accommodate any of these two constraints. Therefore, we have to accept the fact that we shall be dealing with penta-diagonal rather than tridiagonal representations. Despite our long experience in dealing with various basis sets in the J-matrix method, all our attempts to find alternative sets that can support tridiagonal representations while maintaining Hermiticity failed. Consequently, we are faced with the task of reformulating the *J*-matrix method based on penta-diagonal rather than tridiagonal representations. From Eq. (9), we can write the matrix elements of the reference wave operator as follows

$$\langle \phi_m | \mathcal{J} | \phi_n \rangle = -\frac{\lambda^2}{2} \sqrt{\frac{\Gamma(n+1)\Gamma(m+1)}{\Gamma(n+\beta+1)\Gamma(m+\beta+1)}} \int_0^\infty x^{2\alpha-2} e^{-x} L_m^\beta(x) \{....\} dx, \qquad (10)$$

where $\{....\}$ is the curly bracket in Eq. (9). The orthogonality of the Laguerre polynomials dictates that $2\alpha = \beta + 2$. Using this value of $\alpha$ in Eq. (9) and employing the recursion relation of the Laguerre polynomials, we obtain the following action of the reference wave operator on the basis

$$-\frac{2x^2}{\lambda^2} \mathcal{J} \phi_n = \left\{ \nu^2 + \tfrac{1}{4}(\beta^2 - 1) + (\mu^2 - \tfrac{1}{4})(2n+\beta+1)^2 + (\mu^2 + \tfrac{1}{4})\left[2n(n+\beta+1) + \beta+1\right] \right\} \phi_n$$
$$- 2\mu^2 \left[ (2n+\beta)\sqrt{n(n+\beta)}\, \phi_{n-1} + (2n+\beta+2)\sqrt{(n+1)(n+\beta+1)}\, \phi_{n+1} \right] \qquad (11)$$
$$+ (\mu^2 + \tfrac{1}{4})\left[ \sqrt{n(n-1)(n+\beta)(n+\beta-1)}\, \phi_{n-2} + \sqrt{(n+1)(n+2)(n+\beta+1)(n+\beta+2)}\, \phi_{n+2} \right]$$

Therefore, the reference wave equation (1), $\mathcal{J}\psi = 0$, and the expansion of the reference wavefunction (6) as $\psi = \sum_n F_n^\pm \phi_n$, result in the following five-term recursion relation for the expansion coefficients $\{F_n^\pm\}$

$$a_n F_n^\pm + b_{n-1} F_{n-1}^\pm + b_n F_{n+1}^\pm + c_{n-2} F_{n-2}^\pm + c_n F_{n+2}^\pm = 0, \qquad (12)$$

for $n = 2,3,4,...$ and where $\{a_n, b_n, c_n\}$ are the recursion coefficients that read as follows

$$a_n = \nu^2 + \tfrac{1}{4}(\beta^2 - 1) + (\mu^2 - \tfrac{1}{4})(2n+\beta+1)^2 + (\mu^2 + \tfrac{1}{4})\left[2n(n+\beta+1) + \beta+1\right], \qquad (13a)$$
$$b_n = -2\mu^2 (2n+\beta+2)\sqrt{(n+1)(n+\beta+1)}, \qquad (13b)$$
$$c_n = (\mu^2 + \tfrac{1}{4})\sqrt{(n+1)(n+2)(n+\beta+1)(n+\beta+2)}. \qquad (13c)$$

Consequently, the matrix representation of the reference wave operator in the basis $\{\phi_n(r)\}$, which is given by Eq. (10), becomes a symmetric penta-diagonal matrix with the following elements

$$\langle \phi_n | \mathcal{J} | \phi_m \rangle = -\frac{\lambda^2}{2} \left[ a_n \delta_{n,m} + b_{n-1} \delta_{n,m+1} + b_n \delta_{n,m-1} + c_{n-2} \delta_{n,m+2} + c_n \delta_{n,m-2} \right]. \qquad (14)$$

The recursion relation (12) determines all the expansion coefficients $\{F_n^\pm\}_{n=4}^\infty$ starting with the four initial ones $\{F_n^\pm\}_{n=0}^{n=3}$. Therefore, we need to calculate the integral (7) only for $n = 0,1,2,3$. Using the table of integrals on page 91 of Ref. [18], we obtain the following analytic result

$$\int_0^\infty J_{\pm i\nu}(\mu x) e^{-x/2} x^{\frac{\beta-1}{2}+m} dx = \left(\tfrac{1}{2}\sqrt{4\mu^2+1}\right)^{-m-\frac{\beta+1}{2}} \Gamma\left(m+\tfrac{\beta+1}{2}\pm i\nu\right) P_{m+\frac{\beta-1}{2}}^{\mp i\nu}\left(1/\sqrt{4\mu^2+1}\right), \quad (15)$$

where $P_\gamma^\lambda(x)$ is the associated Legendre function of the first kind that could be defined for $0 \le x \le 1$ as follows (see page 167 of Ref. [18])

$$P_\gamma^\lambda(x) = \frac{2^\lambda/\Gamma(1-\lambda)}{(1-x^2)^{\lambda/2}} \, _2F_1\left(\tfrac{1+\gamma-\lambda}{2}, -\tfrac{\gamma+\lambda}{2}; 1-\lambda; 1-x^2\right). \quad (16)$$

Using the series expansion of the Laguerre polynomial, $L_n^\beta(x) = \frac{(\beta+1)_n}{n!} \sum_{m=0}^n \frac{(-n)_m}{(\beta+1)_m} \frac{x^m}{m!}$, in the integral (7), where $(z)_m = z(z+1)(z+2)...(z+m-1) = \frac{\Gamma(z+m)}{\Gamma(z)}$, we can write

$$F_n^\pm(k) = \frac{\pm 1}{\sinh(\nu\pi)} \sqrt{\frac{\mu(\beta+1)_n}{n!\Gamma(\beta+1)}} \sum_{m=0}^n \frac{(-n)_m}{(\beta+1)_m} \frac{1}{m!} \left[ e^{\pm\pi\nu/2} I_m^+(\mu,\nu,\beta) - e^{\mp\pi\nu/2} I_m^-(\mu,\nu,\beta) \right], \quad (17)$$

where $I_m^\pm(\mu,\nu,\beta)$ is the integral (15) and we have used the relation: $A_\pm H_{i\nu}^\pm(x) = \frac{\pm 1}{\sinh(\nu\pi)} \left[ e^{\pm\pi\nu/2} J_{i\nu}(x) - e^{\mp\pi\nu/2} J_{-i\nu}(x) \right]$. This algebraic expression for the integral (7) is more suitable for calculation than direct numerical integration. The four initial values to the five-term recursion (12), $\{F_n^\pm\}_{n=0}^{n=3}$, could now be calculated using (17). For example, the first two are

$$F_0^\pm(k) = \frac{\pm 1}{\sinh(\nu\pi)} \sqrt{\frac{\mu}{\Gamma(\beta+1)}} \left(\sqrt{\mu^2+\tfrac{1}{4}}\right)^{-\frac{\beta+1}{2}}$$
$$\left[ e^{\pm\pi\nu/2} \Gamma\left(\tfrac{\beta+1}{2}+i\nu\right) P_{\frac{\beta-1}{2}}^{-i\nu}\left(1/\sqrt{4\mu^2+1}\right) - e^{\mp\pi\nu/2} \Gamma\left(\tfrac{\beta+1}{2}-i\nu\right) P_{\frac{\beta-1}{2}}^{i\nu}\left(1/\sqrt{4\mu^2+1}\right) \right] \quad (18a)$$

$$F_1^\pm(k) = \sqrt{\beta+1} F_0^\pm(k) + \frac{\mp 1}{\sinh(\nu\pi)} \sqrt{\frac{\mu}{\Gamma(\beta+2)}} \left(\sqrt{\mu^2+\tfrac{1}{4}}\right)^{-\frac{\beta+3}{2}}$$
$$\left[ e^{\pm\pi\nu/2} \Gamma\left(\tfrac{\beta+3}{2}+i\nu\right) P_{\frac{\beta+1}{2}}^{-i\nu}\left(1/\sqrt{4\mu^2+1}\right) - e^{\mp\pi\nu/2} \Gamma\left(\tfrac{\beta+3}{2}-i\nu\right) P_{\frac{\beta+1}{2}}^{i\nu}\left(1/\sqrt{4\mu^2+1}\right) \right] \quad (18b)$$

One can easily verify (analytically or numerically) that the expansion coefficients (17) satisfy the recursion relation (12) for $n = 0$ and $n = 1$ provided that we set $b_{-1} = c_{-1} = c_{-2} \equiv 0$. Therefore, we can calculate $F_2^\pm$ and $F_3^\pm$ using only $F_0^\pm$ and $F_1^\pm$ as follows

$$F_2^\pm = -\frac{1}{c_0}\left(a_0 F_0^\pm + b_0 F_1^\pm\right), \quad (19a)$$

$$F_3^\pm = \frac{b_1}{c_1}\left[\left(\frac{a_0}{c_0} - \frac{b_0}{b_1}\right) F_0^\pm + \left(\frac{b_0}{c_0} - \frac{a_1}{b_1}\right) F_1^\pm\right]. \quad (19b)$$

Thus, if we complement the recursion relation (12) by (19) then the only initial values needed to calculate the complete set $\{F_n^\pm\}$ are $F_0^\pm$ and $F_1^\pm$ given by Eq. (18). In principle, this should have been expected since our reference problem is described by the second order differential equation (1) whose solution requires only two boundary conditions. Now, for a more stable calculation of the expansion coefficients $\{F_n^\pm\}$ we use the recursion relation (12) to compute the ratios $R_n^\pm := F_n^\pm / F_{n-1}^\pm$ starting with the initial ones $\{R_n^\pm\}_{n=1}^{n=3}$. To do that, we divide Eq. (12) by $F_n^\pm$ and obtain

$$R_{n+2}^\pm = -\frac{b_n}{c_n} - \frac{1}{c_n R_{n+1}^\pm}\left[\frac{1}{R_n^\pm}\left(b_{n-1} + \frac{c_{n-2}}{R_{n-1}^\pm}\right) + a_n\right], \tag{20}$$

for $n = 2,3,4,\ldots$. Then, we get $F_n^\pm = R_n^\pm F_{n-1}^\pm = F_0^\pm \prod_{m=1}^n R_m^\pm$.

To test the accuracy and convergence of the *J*-matrix solution of the reference problem, we plot the two sides of equation (6) on the same graph for a given set of parameters $\{\mu,\nu,\beta\}$ and include the first *N* terms of the series for large enough *N*. This is shown in Figure 1 for $N=100$, $N=1000$, and $N=10000$. The anomalies in the inverse square potential manifests itself in the slow convergence of the sum to the exact solution, which we have never encountered during our past experience in dealing with the *J*-matrix method. However, this should have been expected since the extremely rapid oscillation of the exact solution close to the origin can only be matched by similar oscillations in the basis, which can only be achieved by very large degrees of the Laguerre polynomial (i.e., very large basis size). Nonetheless, it is surprising that the *J*-matrix method would produces such equitable asymptotic fit to the exact solution *without* doing any regularization or self-adjoint extension to fix the quantum anomalies in the potential as it is customarily done. Of course, we had to pay the price by struggling to reformulate the theory in penta-diagonal representations, deal with objects that satisfy five-term recursion relations, and cope with slow convergence of the calculation. Figure 2, shows the exact and the series solution in the neighborhood of the origin for $N=10000$.

In the following section, we address the dynamics of the problem by adding the contribution coming from the short-range potential $U(r)$ to the reference Hamiltonian and employing the *J*-matrix recipe to derive the *S*-matrix and phase shift.

### III. THE *S*-MATRIX IN THE PENTA-DIAGONAL REPRESENTATION

Since the range of the potential $U(r)$ is short then we should be able to find a suitable basis set $\{\xi_n(r)\}$ in which the effect of this potential is faithfully represented by its matrix representation in a finite subset, $\{\xi_n\}_{n=0}^{N-1}$, for some large enough integer *N*. Therefore, we divide the representation space into a finite inner subspace of size *N* and a semi-infinite outer subspace. Now since the potential matrix vanishes in the outer subspace, then the natural basis for representations in the outer subspace should be the set $\{\phi_n(r)\}$ given by (5) with $2\alpha = \beta + 2$. Thus, we write the total wavefunction as

$$\Psi(r,E) = \sum_{n=0}^{N-1} p_n(E)\xi_n(r) + \sum_{n=N}^{\infty} q_n(E)\phi_n(r), \qquad (21)$$

where $\{p_n, q_n\}$ are proper expansion coefficients to be determined by solving the total wave equation $(H-E)|\Psi\rangle = (\mathcal{J}+U)|\Psi\rangle = 0$. The infinite matrix representation of the wave operator $(H-E)$ in the total space spanned by the basis $\xi \oplus \phi$ becomes a combination of a finite $N \times N$ matrix representation of $(H-E)$ in the basis $\{\xi_n\}_{n=0}^{N-1}$ augmented by a semi-infinite penta-diagonal symmetric matrix representing $(H_0 - E) = \mathcal{J}$ in the basis $\{\phi_n\}_{n=N}^{\infty}$ as follows:

$$\begin{pmatrix} \times & \times & \times & \times & \times & \times & \times & \times & & & & & & & & & & & & \\ \times & \times & \times & \times & \times & \times & \times & \times & & & & & & & & & & & & \\ \times & \times & \times & \times & \times & \times & \times & \times & & & & & & & & & & & & \\ \times & \times & \times & \times & \times & \times & \times & \times & & & & & & & & & & & & \\ \times & \times & \times & \times & \times & \times & \times & \times & & & & & & & & & & & & \\ \times & \times & \times & \times & \times & \times & \times & \times & & & & & & & & & & & & \\ \times & \times & \times & \times & \times & \times & \times & \times & \otimes & & & & & & & & & & & \\ \times & \times & \times & \times & \times & \times & \times & \times & \otimes & \otimes & & & & & & & & & & \\ & & & & & & & & \otimes & \otimes & \times & \times & \times & & & & & & & \\ & & & & & & & & & \otimes & \times & \times & \times & \times & & & & & & \\ & & & & & & & & & & \times & \times & \times & \times & \times & & & & & \\ & & & & & & & & & & & \times & \times & \times & \times & \times & & & & \\ & & & & & & & & & & & & \times & \times & \times & \times & \times & & & \\ & & & & & & & & & & & & & \times & \times & \times & \times & \times & & \\ & & & & & & & & & & & & & & \times & \times & \times & \times & \times & \\ & & & & & & & & & & & & & & & \times & \times & \times & \times & \times \\ & & & & & & & & & & & & & & & & \times & \times & \times & \times \\ & & & & & & & & & & & & & & & & & \times & \times & \times \end{pmatrix} \begin{pmatrix} p_0 \\ p_1 \\ \times \\ \times \\ \times \\ p_{N-3} \\ p_{N-2} \\ p_{N-1} \\ q_N \\ q_{N+1} \\ q_{N+2} \\ \times \\ \times \\ \times \\ \times \\ \times \\ \times \\ \times \\ \times \\ \times \end{pmatrix} = 0 \quad (22)$$

The finite $N \times N$ matrix block on the top-left corner is the representation of the wave operator $H_0 + U - E$ in the basis $\{\xi_n\}_{n=0}^{N-1}$ whereas the entries outside this block, which are zeros everywhere except for the penta-diagonal tail, represent the reference wave operator $\mathcal{J} = H_0 - E$ in the basis $\{\phi_n\}_{n=N}^{\infty}$. The entries indicated by $\otimes$ constitute couplings between the two subspaces. Each row in (22) gives an equation to be solved for either the $p$'s or the $q$'s. It is evident that rows $N+2$ and beyond result in the following equation

$$\mathcal{J}_{i,i-2}q_{i-2} + \mathcal{J}_{i,i-1}q_{i-1} + \mathcal{J}_{i,i}q_i + \mathcal{J}_{i,i+1}q_{i+1} + \mathcal{J}_{i,i+2}q_{i+2} = 0, \qquad (23)$$

for $i \geq N+2$. Now, the matrix elements $\{\mathcal{J}_{n,m}\}$ given by Eq. (14) show that this is exactly the recursion relation (12). Thus, $\{q_n\}_{n=N}^{\infty}$ must be a linear combination of $F_n^{\pm}$. That is, we can write $q_n = B_+ F_n^+ + B_- F_n^-$, where $B_\pm$ are arbitrary functions of the energy. On the other hand, row $N+1$ of (22) gives the following equation

$$\mathcal{J}_{N+1,N-1} p_{N-1} + \mathcal{J}_{N+1,N} q_N + \mathcal{J}_{N+1,N+1} q_{N+1} + \mathcal{J}_{N+1,N+2} q_{N+2} + \mathcal{J}_{N+1,N+3} q_{N+3} = 0. \tag{24}$$

The matrix elements $\{\mathcal{J}_{n,m}\}$ in Eq. (14) and the recursion relation (12) show that this equation is satisfied if we take $p_{N-1} = q_{N-1} = B_+ F_{N-1}^+ + B_- F_{N-1}^-$. With this result, row $N$ of (22) gives the following equation

$$\mathcal{J}_{N,N-2} p_{N-2} + \mathcal{J}_{N,N-1} q_{N-1} + \mathcal{J}_{N,N} q_N + \mathcal{J}_{N,N+1} q_{N+1} + \mathcal{J}_{N,N+2} q_{N+2} = 0. \tag{25}$$

For the same reason, the solution of this equation is $p_{N-2} = q_{N-2} = B_+ F_{N-2}^+ + B_- F_{N-2}^-$. With these values of $p_{N-1}$ and $p_{N-2}$ and the solution of (23), the infinite matrix equation (22) reduces to the following $N \times N$ finite matrix equation

$$\begin{pmatrix} \times & \times & \times & \times & \times & \times & \times & \times \\ \times & \times & \times & \times & \times & \times & \times & \times \\ \times & \times & \times & \times & \times & \times & \times & \times \\ \times & \times & \times & \times & \times & \times & \times & \times \\ \times & \times & \times & \times & \times & \times & \times & \times \\ \times & \times & \times & \times & \times & \times & \times & \times \\ \times & \times & \times & \times & \times & \times & \times & \times \\ \times & \times & \times & \times & \times & \times & \times & \times \end{pmatrix} \begin{pmatrix} p_0 \\ p_1 \\ \times \\ \times \\ p_{N-4} \\ p_{N-3} \\ q_{N-2} \\ q_{N-1} \end{pmatrix} = \begin{pmatrix} 0 \\ 0 \\ \times \\ \times \\ 0 \\ 0 \\ -\mathcal{J}_{N-2,N} q_N \\ -\mathcal{J}_{N-1,N} q_N - \mathcal{J}_{N-1,N+1} q_{N+1} \end{pmatrix}. \tag{26}$$

The $N \times N$ matrix on the left is the finite representation of the wave operator $H_0 + U - E$ in the basis $\{\xi_n\}_{n=0}^{N-1}$. Let us call the inverse of this finite matrix $\{G_{n,m}\}_{n,m=0}^{N-1}$. That is, in the subspace spanned by $\{\xi_n\}_{n=0}^{N-1}$, $G = (H_0 + U - E)^{-1}$. Then multiplying both sides of Eq. (26) by the matrix $G$, we obtain $N-2$ equations for $\{p_n\}_{n=0}^{N-3}$ in addition to the following two special equations

$$q_{N-2} = -G_{N-2,N-2} \mathcal{J}_{N-2,N} q_N - G_{N-2,N-1} (\mathcal{J}_{N-1,N} q_N + \mathcal{J}_{N-1,N+1} q_{N+1}), \tag{27a}$$

$$q_{N-1} = -G_{N-1,N-2} \mathcal{J}_{N-2,N} q_N - G_{N-1,N-1} (\mathcal{J}_{N-1,N} q_N + \mathcal{J}_{N-1,N+1} q_{N+1}). \tag{27b}$$

This is the asymptotic convergence equation (boundary condition) of the J-matrix method, which states that as $n \to \infty$ (effectively, for all $n \geq N-2$ with $N$ large enough) we have

$$q_n = -G_{n,N-2} \mathcal{J}_{N-2,N} q_N - G_{n,N-1} (\mathcal{J}_{N-1,N} q_N + \mathcal{J}_{N-1,N+1} q_{N+1}), \tag{28a}$$

together with,

$$q_n = B_+ F_n^+ + B_- F_n^-. \tag{28b}$$

Now, what remains is to evaluate $B_\pm$, which is done using the physical boundary condition of scattering and goes as follows. Since the range of the potential $U(r)$ is short, then for a normalized incident flux, the boundary condition at infinity is

$$\lim_{r \to \infty} \Psi(r, E) = \sqrt{\tfrac{2}{\pi}} \left[ e^{+i(kr+\varphi)} - \mathcal{S}(k) e^{-i(kr+\varphi)} \right], \tag{29}$$

where $\mathcal{S}(k)$ is the scattering matrix and $\varphi$ is a real global phase angle. Conservation of probability in elastic scattering dictates that $|\mathcal{S}(k)| = 1$ and we write it as $\mathcal{S}(k) = e^{2i\delta(k)}$ where $\delta(k)$ is the scattering phase shift that depends on the energy and physical parameters. Using the asymptotic formula of the Hankel functions in the reference solution (4) and the expansion (6), we can rewrite the asymptotics (29) as

$$\lim_{r \to \infty} \Psi(r, E) = \lim_{r \to \infty} [\chi_+(r) - \mathcal{S}(k)\chi_-(r)] = \lim_{r \to \infty} \sum_{n=0}^{\infty} \left[ F_n^+(k) - \mathcal{S}(k) F_n^-(k) \right] \phi_n(r). \tag{30}$$

On the other hand, one can show (in fact, we have proved it elsewhere [19]) that the asymptotic equivalence of configuration space and function space states that

$$\lim_{r \to \infty} \sum_{n=0}^{\infty} F_n^\pm(k) \phi_n(r) = \lim_{M \to \infty} \sum_{n=M}^{\infty} F_n^\pm(k) \phi_n(r). \tag{31}$$

Therefore, we can write (29) as follows

$$\lim_{r \to \infty} \Psi(r, E) = \lim_{M \to \infty} \sum_{n=M}^{\infty} \left[ F_n^+(k) - \mathcal{S}(k) F_n^-(k) \right] \phi_n(r). \tag{32}$$

Combining this with the expansion (21) and for large enough integer $N = M$ such that the asymptotic convergence equation (28) holds, we can write for $n \geq N - 2$

$$q_n(E) = F_n^+(k) - \mathcal{S}(k) F_n^-(k), \tag{33}$$

making $B_+ = 1$ and $B_- = -\mathcal{S}(k)$ in (28b). Substituting (33) in (27b), we obtain the following expression for the scattering matrix in the penta-diagonal J-matrix representation

$$\mathcal{S}(k) = T_{N-1} \frac{1 + \left( G_{N-1,N-1} \mathcal{J}_{N-1,N} + G_{N-1,N-2} \mathcal{J}_{N-2,N} \right) R_N^+ + G_{N-1,N-1} \mathcal{J}_{N-1,N+1} R_{N+1}^+ R_N^+}{1 + \left( G_{N-1,N-1} \mathcal{J}_{N-1,N} + G_{N-1,N-2} \mathcal{J}_{N-2,N} \right) R_N^- + G_{N-1,N-1} \mathcal{J}_{N-1,N+1} R_{N+1}^- R_N^-}, \tag{34}$$

where the J-matrix coefficients $T_n(E)$ and $R_n^\pm(E)$ are defined as usual

$$T_n(E) = \frac{F_n^+(k)}{F_n^-(k)} = \frac{C_n(k) + iS_n(k)}{C_n(k) - iS_n(k)}, \quad R_n^\pm(E) = \frac{F_n^\pm(k)}{F_{n-1}^\pm(k)} = \frac{C_n(k) \pm iS_n(k)}{C_{n-1}(k) \pm iS_{n-1}(k)}. \tag{35}$$

It is to be noted that restriction to the tridiagonal representation (where $\mathcal{J}_{N-2,N} = 0$ and $\mathcal{J}_{N-1,N+1} = 0$) reproduces the well-known result of the conventional *J*-matrix

$$S(k) = T_{N-1} \frac{1 + G_{N-1,N-1} \mathcal{J}_{N-1,N} R_N^+}{1 + G_{N-1,N-1} \mathcal{J}_{N-1,N} R_N^-}. \tag{36}$$

It should be clear from Eq. (34) that in the calculation of the *S*-matrix one needs kinematical and dynamical inputs. The kinematic ingredients are the *J*-matrix coefficients $T_n(E)$ and $R_n^\pm(E)$ together with the reference wave operator matrix elements $\mathcal{J}_{n,m}$. These are obtained in Sec. II above. On the other hand, the dynamical ingredients are in the elements of the Green's function $G_{n,m}$ for a given $U(r)$. In Appendix B, we show how to use Gauss quadrature integral approximation to obtain the finite matrix representation of $U(r)$ in the basis set $\{\xi_n(r)\}_{n=0}^{N-1}$ if these are written in a form similar to $\{\phi_n(r)\}$, which is given by Eq. (5). This matrix will be added to the $N \times N$ matrix representation of the reference wave operator, $\langle \xi_n | \mathcal{J} | \xi_m \rangle$, which is different from $\langle \phi_n | \mathcal{J} | \phi_m \rangle$ shown in (14). Consequently, we obtain the finite $N \times N$ matrix representation of the total wave operator in the basis set $\{\xi_n(r)\}_{n=0}^{N-1}$ as $\langle \xi_n | (H_0 + U - E) | \xi_m \rangle$. In Appendix C, we show how to calculate $G_{n,m}$ from this finite square matrix. For computational economy and convenience, we can choose the basis elements for the inner subspace to be the same as that of the outer subspace. That is, we take $\xi_n(r) = \phi_n(r)$.

**IV. CONCLUSION**

The inverse square potential in the supercritical coupling regime is known to be plagued with quantum anomalies. For example, the energy spectrum is abnormal since the potential has no ground state and the energy eigenvalues may assume all negative values that extend to minus infinity. This is despite the fact that the solution corresponding to each of these energy eigenvalues is real and normalizable but each has an infinite number of nodes. Scattering states occur for every positive energy, but the boundary conditions are not sufficient to determine the scattering matrix. This led to a number of renormalization schemes that includes regularization and self-adjoint extension.

In this work, we treated the scattering problem for short-range potentials with inverse square singularity in the supercritical regime using the *J*-matrix method. It is remarkable that we could achieve that without the need for any kind of renormalization. However, we had to pay the price in two ways. Firstly, we had to extend the method to penta-diagonal rather than tridiagonal representations where we build the method on objects that satisfy five-term recursion relations rather than orthogonal polynomials that satisfy three-term recursion relations. Secondly, we had to be contented with slow convergence of the calculation, which we have never encountered during our past experience with the *J*-matrix method.

In the near future, we plan to present several physical examples that demonstrate the viability of the method developed here to handle short-range inverse square singular potentials in the supercritical coupling regime. To test the accuracy of the method, it will be fruitful to include an example that has an exact scattering solution. Unfortunately, all known quantum mechanical systems with short-range inverse square potential are exactly solvable only in the subcritical coupling regime[‡].

## ACKNOWLEDGMENTS

The support by the Saudi Center for Theoretical Physics (SCTP) is highly appreciated. Partial support by King Fahd University of Petroleum and Minerals (KFUPM) is acknowledged.

**APPENDIX A: REFERENCE SOLUTION IN THE OSCILLATOR BASIS**

In this Appendix, we derive the five-term recursion relation and its solution in the *J*-matrix oscillator basis set whose elements are

$$\phi_n(\lambda r) = \sqrt{\tfrac{2\Gamma(n+1)}{\Gamma(n+\beta+1)}} e^{-\lambda^2 r^2/2} (\lambda r)^\alpha L_n^\beta(\lambda^2 r^2). \tag{A1}$$

Consequently, the expansion coefficients $F_n^\pm(k)$ of the two independent solutions (6) of the reference wave equation are evaluated using the orthogonality of the Laguerre polynomials giving the following integral representation

$$F_n^\pm(k) = A_\pm \sqrt{\tfrac{2\mu\Gamma(n+1)}{\Gamma(n+\beta+1)}} \int_0^\infty H_{i\nu}^\pm(\mu x) e^{-x^2/2} x^{2\beta-\alpha+\tfrac{3}{2}} L_n^\beta(x^2) dx. \tag{A2}$$

Applying the reference wave operator (1) on the basis element (A1) and using the differential equation of the Laguerre polynomial and its differential property, we obtain

$$\mathcal{J}\phi_n(x) = -\frac{\lambda^2}{2} e^{-y/2} y^{-1+\alpha/2} \sqrt{\tfrac{2\Gamma(n+1)}{\Gamma(n+\beta+1)}}$$
$$\left\{\left[\nu^2 + \left(\alpha-\tfrac{1}{2}\right)^2 + 4n\left(\alpha-\beta-\tfrac{1}{2}\right) - y\left(4n+2\alpha+1-\mu^2\right) + y^2\right] L_n^\beta(y) - 4\left(\alpha-\beta-\tfrac{1}{2}\right)(n+\beta) L_{n-1}^\beta(y)\right\} \tag{A3}$$

where $y = x^2$. Similar to the Laguerre basis, Hermiticity of the wave operator does not allow for a tridiagonal representation. Now, the matrix elements of the reference wave operator becomes

$$\langle \phi_m | \mathcal{J} | \phi_n \rangle = -\frac{\lambda^2}{2} \sqrt{\tfrac{4\Gamma(n+1)\Gamma(m+1)}{\Gamma(n+\beta+1)\Gamma(m+\beta+1)}} \int_0^\infty y^{\alpha-1} e^{-y} L_m^\beta(y)\{....\} dx, \tag{A4}$$

---

[‡] An example would be the Pöschl-Teller potential, $U(r) = V_0 \sinh^{-2}(\lambda r) + V_1 \cosh^{-2}(\lambda r)$ with $V_0 \leq -\lambda^2/8$. However, this potential is exactly solvable only if $V_0 > -\lambda^2/8$.

−13−

where $\{....\}$ is the curly bracket in Eq. (A3) and $dx = dy/2\sqrt{y}$. The orthogonality of the Laguerre polynomials dictates that $\alpha = \beta + \frac{3}{2}$. Using this value of $\alpha$ and the recursion relation of the Laguerre polynomials into (A3), we obtain the following action of the reference wave operator on the basis

$$-\frac{2x^2}{\lambda^2}\mathcal{J}\phi_n = \left[\nu^2 + (\beta+1)(\mu^2-1) - 2n(n+\beta+1-\mu^2)\right]\phi_n$$
$$-\mu^2\sqrt{n(n+\beta)}\,\phi_{n-1} - \mu^2\sqrt{(n+1)(n+\beta+1)}\,\phi_{n+1} \quad (A5)$$
$$+\sqrt{n(n-1)(n+\beta)(n+\beta-1)}\,\phi_{n-2} + \sqrt{(n+1)(n+2)(n+\beta+1)(n+\beta+2)}\,\phi_{n+2} = 0$$

Therefore, the reference wave equation (1), $\mathcal{J}\psi = 0$, and the expansion of the reference wavefunction (6) as $\psi = \sum_n F_n^\pm \phi_n$, result in the following five-term recursion relation for the expansion coefficients $\{F_n^\pm\}$

$$a_n F_n^\pm + b_{n-1} F_{n-1}^\pm + b_n F_{n+1}^\pm + c_{n-2} F_{n-2}^\pm + c_n F_{n+2}^\pm = 0, \quad (A6)$$

for $n = 2,3,4,...$ and where $\{a_n, b_n, c_n\}$ are the recursion coefficients that read as follows

$$a_n = \nu^2 + (\beta+1)(\mu^2-1) - 2n(n+\beta+1-\mu^2), \quad (A7a)$$
$$b_n = -\mu^2\sqrt{(n+1)(n+\beta+1)}, \quad (A7b)$$
$$c_n = \sqrt{(n+1)(n+2)(n+\beta+1)(n+\beta+2)}. \quad (A7c)$$

Consequently, the matrix representation of the reference wave operator in the basis (A1), which is given by Eq. (A4), becomes a symmetric penta-diagonal matrix with the following elements

$$\langle\phi_n|\mathcal{J}|\phi_m\rangle = -\frac{\lambda^2}{2}\left[a_n\delta_{n,m} + b_{n-1}\delta_{n,m+1} + b_n\delta_{n,m-1} + c_{n-2}\delta_{n,m+2} + c_n\delta_{n,m-2}\right]. \quad (A8)$$

The recursion relation (A6) determines all the expansion coefficients $\{F_n^\pm\}_{n=4}^\infty$ starting with the four initial ones $\{F_n^\pm\}_{n=0}^{n=3}$. Therefore, we need to calculate the integral (A2) only for $n = 0,1,2,3$. Using the table of integrals on page 706 of Ref. [20], we obtain the following analytic result

$$\int_0^\infty J_{\pm i\nu}(\mu x) e^{-x^2/2} x^{\beta+2m} dx = \frac{(\mu/\sqrt{2})^{\pm i\nu} \Gamma\left(m + \frac{1+\beta\pm i\nu}{2}\right)}{(\sqrt{2})^{1-2m-\beta} \Gamma(1\pm i\nu)} {}_1F_1\left(m + \frac{1+\beta\pm i\nu}{2}; 1\pm i\nu; -\mu^2/2\right)$$
$$= \frac{(\mu/\sqrt{2})^{\pm i\nu} \Gamma\left(m + \frac{1+\beta\pm i\nu}{2}\right)}{(\sqrt{2})^{1-2m-\beta} \Gamma(1\pm i\nu)} e^{-\mu^2/2} {}_1F_1\left(-m + \frac{\pm i\nu+1-\beta}{2}; 1\pm i\nu; \mu^2/2\right)$$
(A9)

–14–

where we have also used the identity ${}_1F_1(a;c;-z) = e^{-z}{}_1F_1(c-a;c;z)$. Using this result and the series expansion of the Laguerre polynomial $L_n^\beta(x^2) = \frac{(\beta+1)_n}{n!}\sum_{m=0}^{n}\frac{(-n)_m}{(\beta+1)_m}\frac{x^{2m}}{m!}$ in the integral (A2), we can write

$$F_n^\pm(k) = \frac{\pm 1}{\sinh(\nu\pi)}\sqrt{\frac{2\mu(\beta+1)_n}{n!\Gamma(\beta+1)}}\sum_{m=0}^{n}\frac{(-n)_m}{(\beta+1)_m}\frac{1}{m!}\left[e^{\pm\pi\nu/2}I_m^+(\mu,\nu,\beta) - e^{\mp\pi\nu/2}I_m^-(\mu,\nu,\beta)\right], \quad \text{(A10)}$$

where $I_m^\pm(\mu,\nu,\beta)$ is the integral (A9) and we have used the relation: $A_\pm H_{i\nu}^\pm(x) = \frac{\pm 1}{\sinh(\nu\pi)}\left[e^{\pm\pi\nu/2}J_{i\nu}(x) - e^{\mp\pi\nu/2}J_{-i\nu}(x)\right]$. Using this expression for the expansion coefficients $\{F_n^\pm\}$ one can easily verify (analytically or numerically) that the recursion relation (A6) is satisfied for $n=0$ and $n=1$ provided that we set $b_{-1} = c_{-1} = c_{-2} \equiv 0$. Therefore, we can calculate $F_2^\pm$ and $F_3^\pm$ using only $F_0^\pm$ and $F_1^\pm$ as follows

$$F_2^\pm = -\frac{1}{c_0}\left(a_0 F_0^\pm + b_0 F_1^\pm\right), \quad \text{(A11a)}$$

$$F_3^\pm = \frac{b_1}{c_1}\left[\left(\frac{a_0}{c_0} - \frac{b_0}{b_1}\right)F_0^\pm + \left(\frac{b_0}{c_0} - \frac{a_1}{b_1}\right)F_1^\pm\right]. \quad \text{(A11b)}$$

Consequently, we can calculate the complete set $\{F_n^\pm\}$ using the recursion relation (A6) and only the two initial values $F_0^\pm$ and $F_1^\pm$, which could be evaluated using (A10) as

$$F_0^\pm(k) = \frac{\pm 2^{\beta/2}}{\sinh(\nu\pi)}e^{-\mu^2/2}\sqrt{\frac{\mu}{\Gamma(\beta+1)}}$$

$$\left[e^{\pm\pi\nu/2}\left(\frac{\mu}{\sqrt{2}}\right)^{i\nu}\frac{\Gamma\left(\frac{1+\beta+i\nu}{2}\right)}{\Gamma(1+i\nu)}{}_1F_1\left(\frac{1-\beta+i\nu}{2};1+i\nu;\frac{\mu^2}{2}\right) - e^{\mp\pi\nu/2}\left(\frac{\mu}{\sqrt{2}}\right)^{-i\nu}\frac{\Gamma\left(\frac{1+\beta-i\nu}{2}\right)}{\Gamma(1-i\nu)}{}_1F_1\left(\frac{1-\beta-i\nu}{2};1-i\nu;\frac{\mu^2}{2}\right)\right] \quad \text{(A12a)}$$

$$F_1^\pm(k) = \sqrt{\beta+1}F_0^\pm(k) \mp \frac{2^{1+\beta/2}}{\sinh(\nu\pi)}e^{-\mu^2/2}\sqrt{\frac{\mu}{\Gamma(\beta+2)}}$$

$$\left[e^{\pm\pi\nu/2}\left(\frac{\mu}{\sqrt{2}}\right)^{i\nu}\frac{\Gamma\left(\frac{3+\beta+i\nu}{2}\right)}{\Gamma(1+i\nu)}{}_1F_1\left(-\frac{1+\beta-i\nu}{2};1+i\nu;\frac{\mu^2}{2}\right) - e^{\mp\pi\nu/2}\left(\frac{\mu}{\sqrt{2}}\right)^{-i\nu}\frac{\Gamma\left(\frac{3+\beta-i\nu}{2}\right)}{\Gamma(1-i\nu)}{}_1F_1\left(-\frac{1+\beta+i\nu}{2};1-i\nu;\frac{\mu^2}{2}\right)\right] \quad \text{(A12b)}$$

As in the Laguerre basis, a more stable calculation of the expansion coefficients $\{F_n^\pm\}$ is to compute the ratio $R_n^\pm := F_n^\pm/F_{n-1}^\pm$ using the recursion (A6) and starting with the initial values $\{R_n^\pm\}_{n=1}^{n=3}$. To do that, we divide Eq. (A6) by $F_n^\pm$ and obtain

$$R_{n+2}^\pm = -\frac{b_n}{c_n} - \frac{1}{c_n R_{n+1}^\pm}\left[\frac{1}{R_n^\pm}\left(b_{n-1} + \frac{c_{n-2}}{R_{n-1}^\pm}\right) + a_n\right], \quad \text{(A13)}$$

–15–

for $n = 2, 3, 4, ...$. Then, we get $F_n^\pm = R_n^\pm F_{n-1}^\pm = F_0^\pm \prod_{m=1}^{n} R_m^\pm$. Finally, to obtain the S-matrix in the oscillator basis, we follow the same procedure as we did for the Laguerre basis in Sec. III.

**APPENDIX B: GAUSS QUADRATURE**

All symbols in this Appendix are local and not related to those in the rest of the paper. Let $f(x)$ be a square integrable function with respect to the positive measure $d\mu(x) = \rho(x)dx$ in a real $L^2[x_-, x_+]$ vector space, which is spanned by the complete set of orthonormal basis $\{p_n(x)\}_{n=0}^{\infty}$, where $x \in [x_-, x_+] \subset \mathbb{R}$. Orthogonality is defined as

$$\int_{x_-}^{x_+} \rho(x) p_n(x) p_m(x) dx = \delta_{n,m}. \tag{B1}$$

This suggests that $p_n(x)$ is a polynomial of degree $n$ in $x$ with $\rho(x)$ being its normalized weight function. These polynomials satisfy the following symmetric three-term recursion relation

$$x p_n(x) = a_n p_n(x) + b_{n-1} p_{n-1}(x) + b_n p_{n+1}(x), \tag{B2}$$

for $n = 0, 1, 2,...$ and where the "recursion coefficients" $\{a_n, b_n\}$ are real constants such that $b_n \neq 0$ for all $n \geq 0$ and $b_{-1} \equiv 0$. This recursion gives all the polynomials of any degree starting with the initial seed $p_0(x) = 1$. Associated with this space is an infinite dimensional real tridiagonal symmetric matrix $J$ whose elements are

$$J_{n,m} = a_n \delta_{n,m} + b_{n-1} \delta_{n,m+1} + b_n \delta_{n,m-1} = \int_{x_-}^{x_+} x \rho(x) p_n(x) p_m(x) dx = \delta_{n,m}. \tag{B3}$$

For numerical computations, however, the space is truncated to a finite $N$-dimensional space spanned by $\{p_n(x)\}_{n=0}^{N-1}$. The tridiagonal matrix (B3) becomes a finite $N \times N$ matrix $J$. The distinct real eigenvalues of $J$, which we designate as the set $\{\varepsilon_n\}_{n=0}^{N-1}$, are the zeros of the polynomial $p_N(x)$; i.e. $p_N(\varepsilon_n) = 0$. Let $\{\Lambda_{m,n}\}_{m=0}^{N-1}$ be the normalized eigenvector of $J$ associated with the eigenvalue $\varepsilon_n$. In this setting, Gauss quadrature integral approximation states that

$$\int_{x_-}^{x_+} \rho(x) f(x) dx \cong \sum_{n=0}^{N-1} \omega_n f(\varepsilon_n), \tag{B4}$$

where the "numerical weights" $\omega_n = \Lambda_{0,n}^2$. We can also write these numerical weights in terms of $\{\varepsilon_n\}_{n=0}^{N-1}$ and another set of eigenvalues $\{\hat{\varepsilon}_m\}_{m=0}^{N-2}$ as

$$\omega_n = \frac{\prod_{\substack{m=0}}^{N-2}(\varepsilon_n - \hat{\varepsilon}_m)}{\prod_{\substack{k=0 \\ k \neq n}}^{N-1}(\varepsilon_n - \varepsilon_k)}, \tag{B5}$$

where $\{\hat{\varepsilon}_m\}_{m=0}^{N-2}$ is the eigenvalues of a submatrix of $J$ obtained by deleting the first (zeroth) row and first column. These eigenvalues interweave as $\varepsilon_0 < \hat{\varepsilon}_0 < \varepsilon_1 < \hat{\varepsilon}_1 < \varepsilon_2 ....\hat{\varepsilon}_{N-2} < \varepsilon_{N-1}$. The integral approximation (B4) becomes exact if $f(x)$ is a polynomial in $x$ of a degree less than or equal to $2N-1$. If, instead of the integral (B4), we have $\int_{x_-}^{x_+} f(x) dx$ then this could be approximated as follows

$$\int_{x_-}^{x_+} f(x) dx = \int_{x_-}^{x_+} \rho(x) \frac{f(x)}{\rho(x)} dx \cong \sum_{n=0}^{N-1} \omega_n \frac{f(\varepsilon_n)}{\rho(\varepsilon_n)} = \sum_{n=0}^{N-1} \tilde{\omega}_n f(\varepsilon_n), \tag{B6}$$

where $\tilde{\omega}_n$ is named the "derivative weight" and $\tilde{\omega}_n = \omega_n / \rho(\varepsilon_n)$.

Very often, we encounter integrals that represent matrix elements of functions, similar to the matrix elements of a potential function, and of the form

$$f_{n,m} = \int_{x_-}^{x_+} \rho(x) p_n(x) f(x) p_m(x) dx. \tag{B7}$$

Using Eq. (B4), this integral is approximated as follows

$$f_{n,m} \cong \sum_{k=0}^{N-1} \omega_k p_n(\varepsilon_k) f(\varepsilon_k) p_m(\varepsilon_k). \tag{B8}$$

Now, it is well established that $p_n(\varepsilon_k) = \Lambda_{n,k}/\Lambda_{0,k}$ for $n,k = 0,1,2,..,N-1$. Therefore, with $\omega_k = \Lambda_{0,k}^2$ this could be rewritten in matrix form as

$$f_{n,m} \cong \sum_{k=0}^{N-1} \Lambda_{n,k} f(\varepsilon_k) \Lambda_{m,k} = \left(\Lambda F \Lambda^T\right)_{n,m}, \tag{B9}$$

where $F$ is a diagonal matrix with elements $F_{n,k} = f(\varepsilon_k)\delta_{n,k}$. Therefore, to obtain an approximate evaluation of integrals using Gauss quadrature associated with orthogonal polynomials satisfying (B1) and (B2), one needs only the tridiagonal symmetric matrix $J$, which is constructed using the recursion coefficients $\{a_n, b_n\}$, and possibly the weight function $\rho(x)$ for integrals of the type (B6).

**APPENDIX C: THE FINITE GREEN'S FUNCTION IN ANY $L^2$ BASIS**

All symbols in this Appendix are local and not related to those in the rest of the paper. Let $\{\psi_n\}_{n=0}^{\infty}$ be a complete $L^2$ basis (not necessarily orthogonal) in the configuration space that supports a Hermitian representation for the wave operator, $H - z$, where $H$ is the Hamiltonian and $z$ is a real number. The orthogonal conjugate space is spanned by $\{\bar{\psi}_n\}_{n=0}^{\infty}$, where $\langle \bar{\psi}_n | \psi_m \rangle = \langle \psi_n | \bar{\psi}_m \rangle = \delta_{nm}$. The Green's function $G(z)$ is formally defined by $G(z)(H-z) = 1$. Now, since the wave operator matrix is defined in the basis $\{\psi_n\}$, then using the completeness of the basis $\sum_k |\bar{\psi}_k\rangle\langle\psi_k| = 1$, we can write the matrix elements of this defining equation as: $\sum_k \langle \bar{\psi}_n | G | \bar{\psi}_k \rangle \langle \psi_k | (H-z) | \psi_m \rangle = \delta_{n,m}$. Consequently, the matrix representation of the Green's function $G(z)$ is properly defined in the conjugate basis $\{\bar{\psi}_n\}$ and thus given as

$$G_{nm}(z) = \langle \bar{\psi}_n | (H-z)^{-1} | \bar{\psi}_m \rangle, \tag{C1}$$

On the other hand, manipulation of Green functions involving inverse of operators is carried out most appropriately in an orthogonal basis $\{\chi_n\}_{n=0}^{\infty}$. We start by writing the eigenvalue problem in this basis as

$$H|\chi_n\rangle = \varepsilon_n |\chi_n\rangle \quad , n = 0, 1, .., N-1 \tag{C2}$$

From now on, we work in a finite subspace of dimension $N$. Since the matrix representations of the relevant operators are given in the basis $\{\psi_n\}$ rather than $\{\chi_n\}$, then we project from left by $\langle \psi_m |$ and rewrite this equation in the following form

$$\sum_{k=0}^{N-1} \langle \psi_m | H | \psi_k \rangle \langle \bar{\psi}_k | \chi_n \rangle = \varepsilon_n \sum_{k=0}^{N-1} \langle \psi_m | \psi_k \rangle \langle \bar{\psi}_k | \chi_n \rangle \quad ; n, m = 0, 1, .., N-1 \tag{C3}$$

where we have used the completeness property of the basis in the finite $N$ dimensional subspace. That is,

$$\sum_{k=0}^{N-1} |\psi_k\rangle\langle\bar{\psi}_k| = \sum_{k=0}^{N-1} |\bar{\psi}_k\rangle\langle\psi_k| = I, \tag{C4}$$

where $I$ is the $N \times N$ identity matrix. In matrix notation, equation (C3) reads

$$\sum_{k=0}^{N-1} H_{mk} \zeta_k^n = \varepsilon_n \sum_{k=0}^{N-1} \Omega_{mk} \zeta_k^n \quad ; n, m = 0, 1, .., N-1 \tag{C5}$$

where $\zeta_k^n = \langle \bar{\psi}_k | \chi_n \rangle$ and $\Omega_{nm}$ is the overlap matrix element $\langle \psi_n | \psi_m \rangle$. Thus, $\{\zeta_k^n\}_{k=0}^{N-1}$ is the generalized normalized eigenvector associated with the eigenvalue $\varepsilon_n$. This is so because Eq. (C5) could be viewed as the following generalized matrix eigenvalue equation,

$$H|\zeta^n\rangle = \varepsilon_n \Omega |\zeta^n\rangle. \tag{C6}$$

−18−

The definition $\zeta_k^n = \langle \bar{\psi}_k | \chi_n \rangle$ could also be used to write the orthogonal basis as $|\chi_n\rangle = \sum_m \zeta_m^n |\psi_m\rangle$. Let us define the eigenvectors matrix $\Gamma_{nm} \equiv \zeta_n^m = \langle \bar{\psi}_n | \chi_m \rangle$. Then Eq. (C5) reads $(H\Gamma)_{mn} = \varepsilon_n (\Omega\Gamma)_{mn}$ which when multiplied from left by $\Gamma^{\mathsf{T}}$, where $\Gamma^{\mathsf{T}}_{nm} = \langle \chi_n | \bar{\psi}_m \rangle$, gives

$$(\Gamma^{\mathsf{T}} H \Gamma)_{mn} = \varepsilon_n (\Gamma^{\mathsf{T}} \Omega \Gamma)_{mn} \quad ; n, m = 0, 1, .., N-1. \tag{C7}$$

The matrix $\Gamma$ *simultaneously* diagonalizes $H$ and $\Omega$. That is,

$$(\Gamma^{\mathsf{T}} H \Gamma)_{nm} = \eta_n \delta_{nm}, \quad \text{and} \quad (\Gamma^{\mathsf{T}} \Omega \Gamma)_{nm} = \tau_n \delta_{nm}. \tag{C8}$$

Henceforth, we deduce that $\varepsilon_n = \eta_n / \tau_n$ and Eq. (C1) could be written as

$$\begin{aligned}
G_{nm}(z) &= \sum_{i,j,k,l=0}^{N-1} \langle \bar{\psi}_n | \chi_i \rangle \langle \chi_i | \bar{\psi}_k \rangle \langle \psi_k | (H-z)^{-1} | \psi_l \rangle \langle \bar{\psi}_l | \chi_j \rangle \langle \chi_j | \bar{\psi}_m \rangle \\
&= \sum_{i,j,k,l=0}^{N-1} \Gamma_{ni} \left\{ \Gamma^{\mathsf{T}}_{ik} \left[ (H - z\Omega)^{-1} \right]_{kl} \Gamma_{lj} \right\} \Gamma^{\mathsf{T}}_{jm}
\end{aligned} \tag{C9}$$

Now,

$$\sum_{k,l=0}^{N-1} \Gamma^{\mathsf{T}}_{ik} \left[ (H - z\Omega)^{-1} \right]_{kl} \Gamma_{lj} = \frac{\delta_{ij}}{\eta_i - z\tau_i} = \frac{1}{\tau_i} \frac{\delta_{ij}}{\varepsilon_i - z}. \tag{C10}$$

Therefore, we finally obtain

$$G_{nm}(z) = \sum_{i=0}^{N-1} \frac{\Gamma_{ni} \Gamma_{mi}}{\eta_i - z\tau_i} = \sum_{i=0}^{N-1} \frac{1}{\tau_i} \frac{\Gamma_{ni} \Gamma_{mi}}{\varepsilon_i - z}. \tag{C11}$$

For orthogonal basis, that is $\psi_n = \bar{\psi}_n$, the overlap matrix is just the identity matrix $I$, hence $\tau_i = 1$, $\eta_i = \varepsilon_i$. In this orthogonal basis, we can write

$$G_{nm}(z) = \sum_{i=0}^{N-1} \frac{\Gamma_{ni} \Gamma_{mi}}{\varepsilon_i - z} \qquad \underline{\text{Orthogonal Basis}} \tag{C12}$$

**An alternative expression for $G_{nm}(z)$:**
Numerically, it is preferred to work with eigenvalues of matrices rather than their eigenvectors. Here we give an alternative formula to (C11) whereby only eigenvalues are involved. Let $\overset{nm}{H}$ and $\overset{nm}{\Omega}$ be submatrices of $H$ and $\Omega$ obtained by deleting the $n^{\text{th}}$ row and $m^{\text{th}}$ column. The eigenvalue and generalized eigenvalue equations in the truncated space, which parallel equations (C2) and (C6), are

$$\overset{nm}{H} | \tilde{\chi}_k \rangle = \tilde{\varepsilon}_k | \tilde{\chi}_k \rangle, \tag{C13}$$

–19–

$$\overset{nm}{H}|\tilde{\zeta}^k\rangle = \overset{nm}{\varepsilon_k} \overset{nm}{\Omega}|\tilde{\zeta}^k\rangle, \tag{C14}$$

where $k = 0, 1, .., N-2$. Similarly, we define the corresponding eigenvectors matrix $\tilde{\Gamma}_{ij} \equiv \tilde{\zeta}_i^j = \langle \overset{nm}{\bar{\psi}_i} | \tilde{\chi}_j \rangle$ which simultaneously diagonalizes $\overset{nm}{H}$ and $\overset{nm}{\Omega}$

$$\left( \tilde{\Gamma}^\intercal \overset{nm}{H} \tilde{\Gamma} \right)_{ij} = \overset{nm}{\eta_i} \delta_{ij} \quad \text{and} \quad \left( \tilde{\Gamma}^\intercal \overset{nm}{\Omega} \tilde{\Gamma} \right)_{ij} = \overset{nm}{\tau_i} \delta_{ij} \tag{C15}$$

and can also write $\overset{nm}{\varepsilon_k} = \overset{nm}{\eta_k} / \overset{nm}{\tau_k}$. Then, it could be shown that the following is an alternative but equivalent form for $G_{nm}(z)$

$$G_{nm}(z) = (-1)^{n+m} \frac{|\overset{nm}{\Omega}|}{|\Omega|} \frac{\prod_{j=0}^{N-2} \overset{nm}{\varepsilon_i} - z}{\prod_{j=0}^{N-1} \varepsilon_j - z} = (-1)^{n+m} \left( \frac{\prod_{i=0}^{N-2} \overset{nm}{\xi_i}}{\prod_{j=0}^{N-1} \xi_j} \right) \frac{\prod_{i=0}^{N-2} \overset{nm}{\varepsilon_i} - z}{\prod_{j=0}^{N-1} \varepsilon_j - z} \tag{C16}$$

where $\{\xi_n\}_{n=0}^{N-1}$ and $\{\overset{nm}{\xi_k}\}_{k=0}^{N-2}$ are the eigenvalues of the overlap matrices $\Omega$ and $\overset{nm}{\Omega}$, respectively. In orthogonal bases, equation (C16) could be written as

$$G_{nm}(z) = (-1)^{n+m} \frac{\prod_{i=0}^{N-2} \overset{nm}{\varepsilon_i} - z}{\prod_{j=0}^{N-1} \varepsilon_j - z} \qquad \text{Orthogonal Basis} \tag{C17}$$

A by-product of formulas (C12) and (C17) for orthogonal bases is the following important relation

$$\Gamma_{nk} \Gamma_{mk} = (-1)^{n+m} \frac{\prod_{i=0}^{N-2} \overset{nm}{\varepsilon_i} - \varepsilon_k}{\prod_{\substack{j=0 \\ j \neq k}}^{N-1} \varepsilon_j - \varepsilon_k} \qquad \text{Orthogonal Basis,} \tag{C18}$$

which is obtained by evaluating (C12) and (C17) at $z = \varepsilon_k$. A special case of this relation (for $n = m$) is a way to calculate the normalized eigenvectors of a square Hermitian matrix in terms of its eigenvalues

$$\Gamma_{nk}^2 = \frac{\prod_{i=0}^{N-2} \overset{nn}{\varepsilon_i} - \varepsilon_k}{\prod_{\substack{j=0 \\ j \neq k}}^{N-1} \varepsilon_j - \varepsilon_k} \qquad \text{Orthogonal Basis.} \tag{C19}$$

This formula has been rediscovered recently by D. M. Mitnik and S. A. H. Mitnik [21]. However, we have been utilizing it since the early inception of the *J*-matrix method. A generalization of (C18) to non-orthogonal basis is obtained by combining (C11) and (C16) where we obtain the following in a generalized (not necessarily orthogonal) basis

$$\Gamma_{nk}\Gamma_{mk} = (-1)^{n+m}\tau_k \frac{\left|\Omega\right|^{nm} \prod_{i=0}^{N-2} \overset{nm}{\varepsilon_i - \varepsilon_k}}{\left|\Omega\right| \prod_{\substack{j=0 \\ j \neq k}}^{N-1} \varepsilon_j - \varepsilon_k} \tag{C20}$$

It should be noted that the terms "orthogonal" and "non-orthogonal" in the language of bases used above correspond (in the language of matrix equations) to the terms "eigenvalue equation" and "generalized eigenvalue equation", respectively.

[14] H. E. Camblong, L. N. Epele, H. Fanchiotti, C. A. García Canal, and C. R. Ordóñez, *On the inequivalence of renormalization and self-adjoint extensions for quantum singular interactions*, Phys. Lett. A **364**, 458 (2007)
[15] A. D. Alhaidari, *Renormalization of the strongly attractive inverse square potential: Taming the singularity*, Found. Phys. **44**, 1049 (2014)
[16] G. B. Arfken and H. J. Weber, *Mathematical Methods for Physicists*, 6th ed. (Elsevier, Academic Press, 2005)
[17] F. W. J. Olver, *Asymptotics and Special Functions* (Academic Press, New York, 1974)
[18] W. Magnus, F. Oberhettinger, and R. P. Soni, *Formulas and Theorems for the Special Functions of Mathematical Physics*, 3rd ed. (Springer, Berlin, 1966)
[19] A. D. Alhaidari, *On the asymptotic solutions of the scattering problem*, J. Phys. A **41**, 175201 (2008)
[20] I. S. Gradshteyn and I. M. Ryzhik, *Table of Integrals, Series, and Products*, Editors: A. Jeffrey and D. Zwillinger, 7th ed. (Elsevier and Academic, 2007)
[21] D. M. Mitnik and S. A. H. Mitnik, *Wavefunctions from energies: Applications in simple potentials*, J. Math. Phys. **61**, 062101 (2020) and references therein

**FIGURE CAPTIONS:**

**FIG. 1**: Plot of the reference solution $\chi_+(r)$ as given by Eq. (4) in red and the *J*-matrix solution as given by the sum on the right side of Eq. (6) in blue. The sum in Eq. (6) is truncated up to $n = N$ and we took $\mu = 2.0$, $\nu = 3$, and $\beta = 4$. The real part of the solution is on the left column of the figure and the imaginary part is on the right. The three sizes of the basis set are taken as $N = 100, 1000, 10000$ from top to bottom row. The horizontal axis is the radial coordinate *r* in units of $\lambda^{-1}$. We did not plot $\chi_-(r)$ since $\text{Re}\,\chi_-(r) = \text{Re}\,\chi_+(r)$ and $\text{Im}\,\chi_-(r) = -\text{Im}\,\chi_+(r)$.

**FIG. 2**: Reproduction of Figure 1 for $N = 10000$ but closer to the origin. The bottom row shows poorer convergence and accuracy as we come closer to the origin.

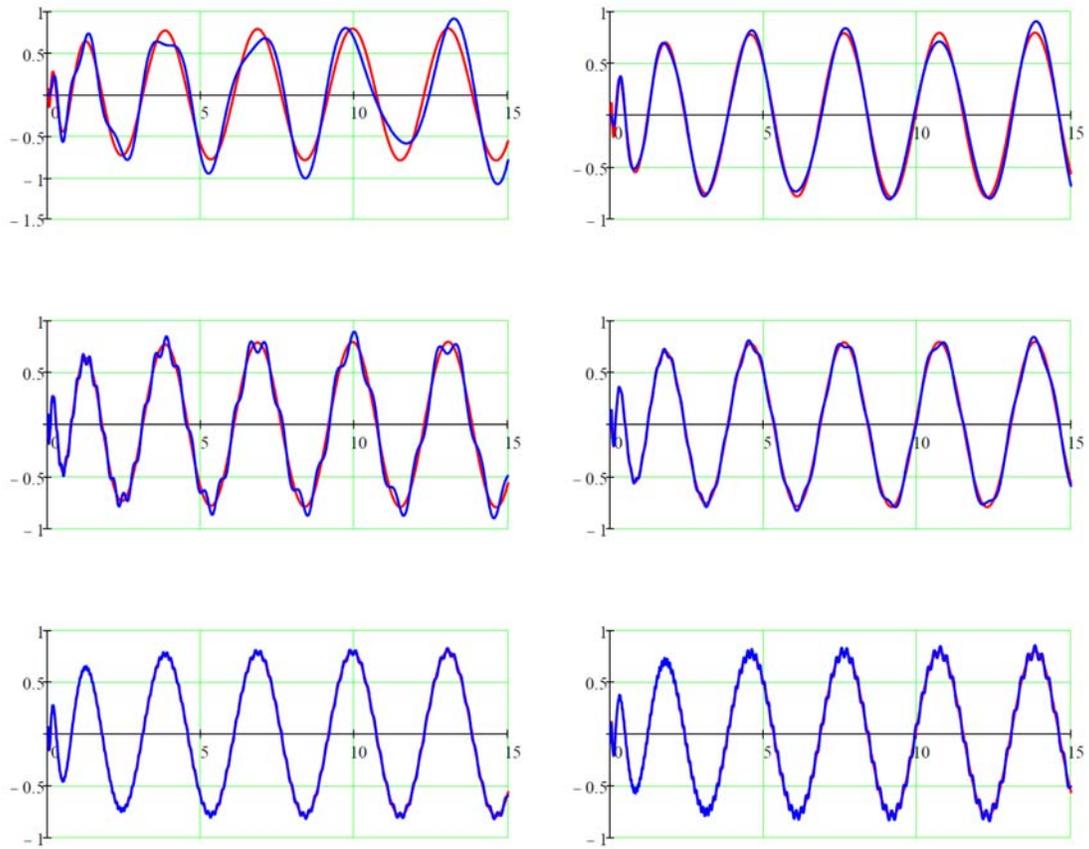

**FIG. 1**

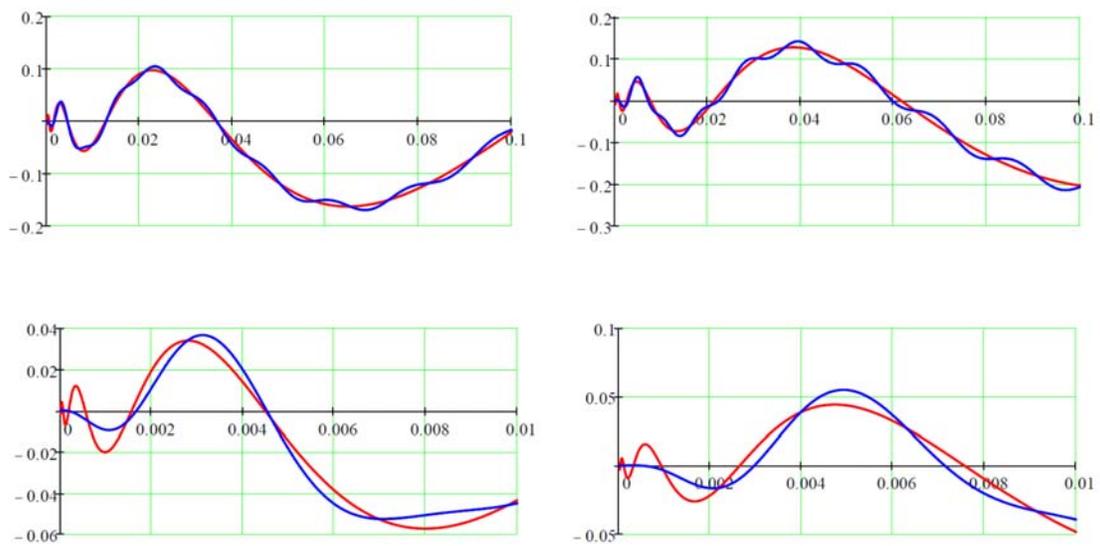

**FIG. 2**